\documentclass[aps,prc,twocolumn,groupedaddress]{revtex4}

\usepackage{graphicx}
\usepackage{longtable}
\def\vc#1{\mbox{\boldmath $#1$}}
\usepackage{subfigure}
\begin{document}
%\setlength{\baselineskip}{25pt}

%\preprint{}

\title{Inelastic form factors to alpha particle condensate states \\ in $^{12}$C and $^{16}$O: 
what can we learn ?}


\author{Y.~Funaki$^1$, A.~Tohsaki$^2$, H.~Horiuchi$^1$, P.~Schuck$^3$,
 and G.~R\"opke$^4$}
%\thanks{}
\affiliation{$^1$ Department of Physics, Kyoto University, Kyoto 606-8502, Japan \\
$^2$ Suzuki Corporation, 46-22 Kamishima-cho Kadoma, Osaka 571-0071 Japan \\
$^3$ Institut de Physique Nucl\'eaire, 91406 Orsay Cedex, France \\
$^4$ Institut f\"ur Physik, Universit\"at Rostock, D-18051 Rostock, Germany
}
\date{\today}

\begin{abstract}
In order to discuss the spatial extention of the $0_2^+$ state of $^{12}$C (Hoyle state), we analyze the inelastic form factor of electron scattering to the Hoyle state, which our $3\alpha$ condensate wave function reproduces very well like previous $3\alpha$ RGM/GCM models. The analysis is made by varying the size of the Hoyle state artificially. As a result, we find that only the maximum value of the form factor sensitively depends on its size, while the positions of maximum and minimum are almost unchanged. This size dependence is found to come from a characteristic feature of the transition density from the ground state to the Hoyle state. We further show the theoretical predictions of the inelastic form factor to the $2_2^+$ state of $^{12}$C, which was recently observed above the Hoyle state, and of the inelastic form factor to the calculated $0_3^+$ state of $^{16}$O, which was conjectured to correspond to the $4\alpha$ condensed state in previous theoretical work by the present authors. 
\end{abstract}

%\keywords{}


\maketitle

 In this short communication we want to report on our calculation of the elastic and inelastic form factors for ground and Hoyle states in $^{12}$C. We also make predictions for the inelastic form factors, $0_1^+ \rightarrow 2_2^+$ in $^{12}$C and $0^+_1 \rightarrow \alpha$-condensate state in $^{16}$O. 

 As we have shown previously, our alpha particle wave function, published in \cite{THSR,funaki}, very nicely reproduces on the one hand some experimental data in $^{12}$C and on the other hand it is in close agreement with calculated results of Kamimura {\it et al.}~\cite{funaki,kamimura}. It is therefore not so surprising that our wave function also reproduces very well the elastic and inelastic $(0_1^+ \rightarrow 0_2^+)$ form factors in $^{12}$C as this was the case in \cite{kamimura}. This gives quite definite support to our interpretation of the Hoyle state as being a condensate of almost independent alpha particles~\cite{yamada,matsumura}.

The form factor is obtained by performing the Fourier transformation of the transition density as follows,
\begin{equation}
|F(q)|^2=\frac{4\pi}{12^2}\Big| \int_0^{\infty} \rho^{(J)}_{J,0_1}(r)j_J(qr)r^2 dr 
\Big|^2 \exp \Big(-\frac{1}{2}a_p^2 q^2 \Big). \label{eq:1}
\end{equation}
Here $a_p^2=0.43$ fm$^2$ is taken as the finite proton size, 
which is the same as adopted in Ref.~\cite{kamimura} and 
$j_J(qr)$ is the $J$-th order spherical Bessel function. 
The transition density $\rho^{(J)}_{J,0_1}(r)$ is given as,
\begin{equation}
\rho^{(J)}_{J,0_1}(r)=\langle \Psi^{JM}_{\lambda=k} 
|\sum_{i=1}^{12}\delta(\vc{r}-\vc{r}_i)| \Psi^{J=0}_{\lambda=1}  
\rangle / Y^\ast_{JM}({\widehat {\vc r}}), \label{eq:2}
\end{equation}
where the ground and Hoyle states can be obtained by solving the following Hill-Wheeler equation,
\begin{equation}
\sum_{\vc{\beta^\prime}}
\big\langle {\widehat \Phi}^{N, J=0}_{3\alpha}(\vc{\beta}) \big| (H-E) \big| {\widehat \Phi}^{N, J=0}_{3\alpha}(\vc{\beta^\prime}) \big\rangle f^{J=0}_\lambda (\vc{\beta^\prime}) =0 .  \label{eq:3}
\end{equation}
\begin{equation}
\Psi^{J=0}_\lambda = \sum_{\vc{\beta}} f^{J=0}_\lambda (\vc{\beta}) 
{\widehat \Phi}^{N, J=0}_{3\alpha}(\vc{\beta}). \label{eq:4}
\end{equation}
We here use the same notation for the alpha condensate wave function,
${\widehat \Phi}^{N, J=0}_{3\alpha}(\vc{\beta})$ as was done in Ref.~\cite{funaki}. The Hamiltonian $H$ is the same as used in Ref.~\cite{funaki,kamimura}. The ground and Hoyle states correspond to the cases of $\lambda=1$ and $\lambda=2$ in Eq.~(\ref{eq:4}), respectively.

\begin{figure}[hbp]
\begin{center}
\subfigure[]{
\includegraphics[width=6cm,height=6cm]{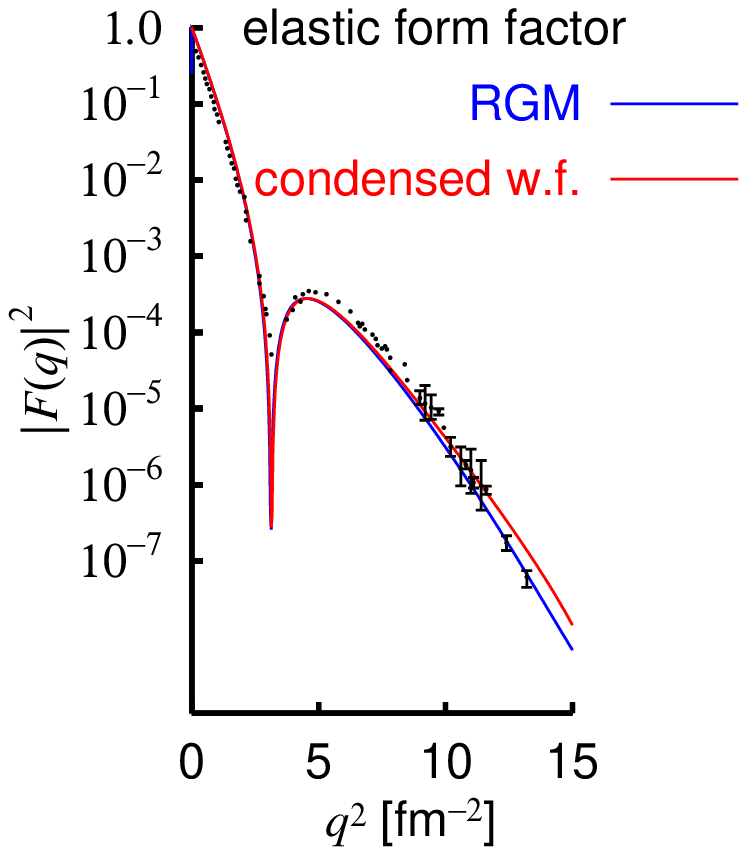}}
\subfigure[]{
\includegraphics[width=6cm,height=6cm]{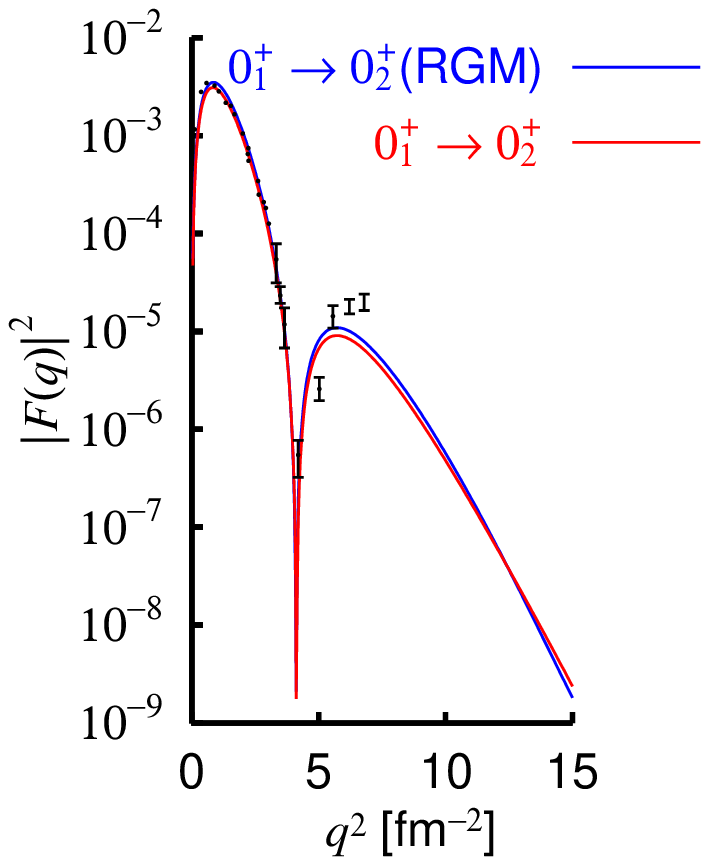}}
\end{center}
\caption{
(Color online) (a): Experimental values of elastic form factor in $^{12}$C are compared with our values obtained by solving the Hill-Wheeler equation Eq.~(\ref{eq:3}) for $\Psi^{J=0}_{\lambda=1}$. The result given in Ref.~\cite{kamimura} using resonating group method (RGM) is also shown. (b): Experimental values of inelastic form factor in $^{12}$C to the Hoyle state are compared with our values obtained by using the Hoyle state wave function $\Psi^{J=0}_{\lambda=2}$ and those given in Ref.~\cite{kamimura} (RGM). The experimental values are taken from Ref~\cite{fmfct_exp}.}\label{fig:1}
\end{figure}

\begin{figure}[htbp]
\begin{center}
\subfigure[]{
\includegraphics[width=6cm,height=6cm]{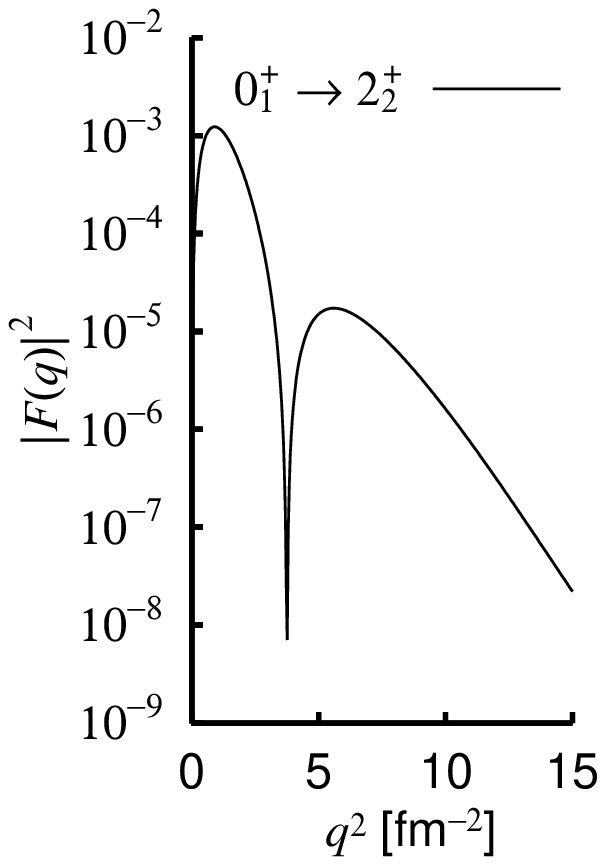}}
\subfigure[]{
\includegraphics[width=6cm,height=6cm]{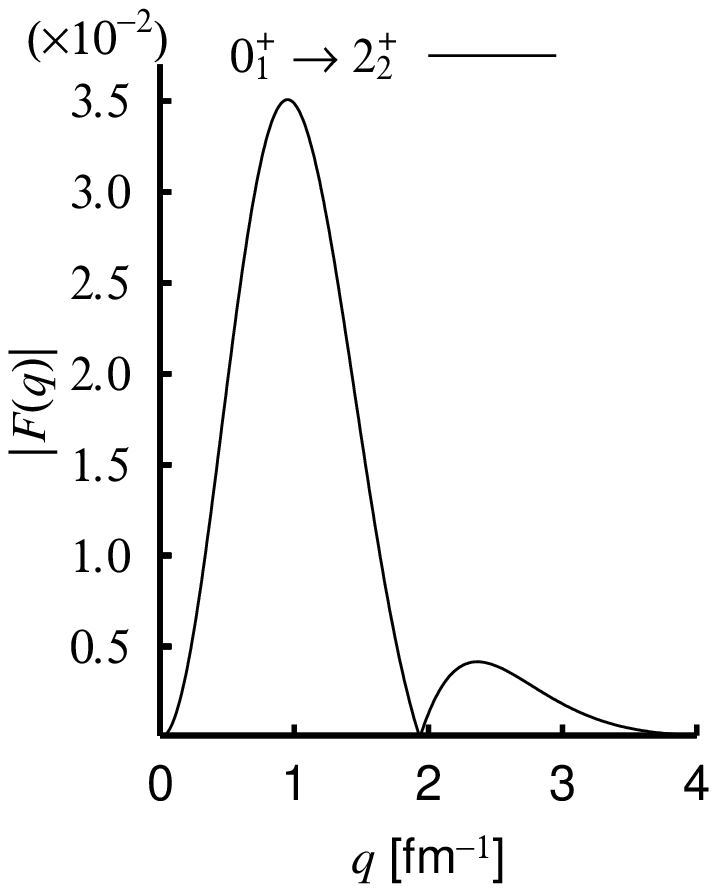}}
\end{center}
\caption{
(a): Theoretical prediction of inelastic form factor to the $2_2^+$ state of $^{12}$C using the wave function of \cite{funaki2}. (b): The form factor shown in (a) is plotted as a function of $q$ in linear scale for ordinate.}\label{fig:1_5}
\end{figure}

 Our results are shown in FIG.~\ref{fig:1} and we give our numerical values in Table~\ref{tab:1}. Reflecting the fact that our wave functions of the ground state and the $0_2^+$ state are almost equivalent to those given in Ref.~\cite{kamimura} using resonating group method (RGM), our elastic and inelastic $(0_1^+ \rightarrow 0_2^+)$ form factors almost completely agree with those given in Ref.~\cite{kamimura}. In FIG.~\ref{fig:1_5}, we predict the inelastic form factor to the $2_2^+$ state which is obtained in Ref.~\cite{funaki2} by using the $3\alpha$ condensate wave functions. This state was recently observed at $2.6 \pm 0.3 $ MeV above the three alpha threshold~\cite{itoh}, though the exsistence of this state has been suggested for a long time from the theoretical point of view~\cite{carbon}. Recently this state was carefully investigated by the present authors~\cite{funaki2} showing that the $2_2^+$ state is intimately related to the $0_2^+$ state which is interpreted as the $3\alpha$ Bose-condensate state. 

 We now make a study of the sensitivity of the inelastic form factor with respect to some theoretical ingredients of our theory. A quantity of prime interest is the spatial extention of the Hoyle state which is predicted from our studies to have a volume $3$ to $4$ times as large as the one of the ground state of $^{12}$C. We therefore repeated the calculation of the inelastic form factor in varying the size parameter of the Hoyle state. The calculation can be done as shown in Ref.~\cite{funaki} by adopting as the Hoyle state the wave function, $\Psi_\perp(\vc{\beta})\equiv{\widehat P}_\perp {\widehat \Phi}^{N, J=0}_{3\alpha}(\vc{\beta})$, where ${\widehat P}_\perp$ is defined as the projection operator onto the orthogonal space to the ground state:

\begin{equation}
{\widehat P}_\perp \equiv 1-| \Psi^{J=0}_{\lambda=1} \rangle \langle \Psi^{J=0}_{\lambda=1} |.
\end{equation}

\begin{table*}[htbp]
\begin{center}
\caption{
Numerical values of elastic (upper row) and inelastic (lower row) form factors in $^{12}$C.}
\begin{tabular}{ccccccccccccc}
\hline
$q$ [fm$^{-1}$] & 0.25 & 0.50 & 0.75 & 1.00 & 1.25 & 1.50 & 1.75 & 2.00 & 2.25 & 2.50 & 2.75 & 3.00  \\
\hline\hline
$|F(q)_{0_1^+ \rightarrow 0_1^+}|$$(\times10^{-1})$ & $9.4$ & $7.7$ & $5.5$ & $3.3$ & $1.6$ & $0.56$ & $0.030$ & $0.15$ & $0.16$ & $0.11$ & $0.065$ & $0.033$  \\
$|F(q)_{0_1^+ \rightarrow 0_2^+}|$$(\times10^{-2})$ & 0.98 & 3.3 & 5.2 & 5.5 & 4.3 & 2.5 & 0.96 & 0.069 & 0.27 & 0.29 & 0.20 & 0.11  \\
\hline
\end{tabular}
\label{tab:1}
\end{center}
\end{table*}

\begin{figure}[htbp]
\begin{center}
\subfigure[]{
\includegraphics[width=6cm,height=6cm]{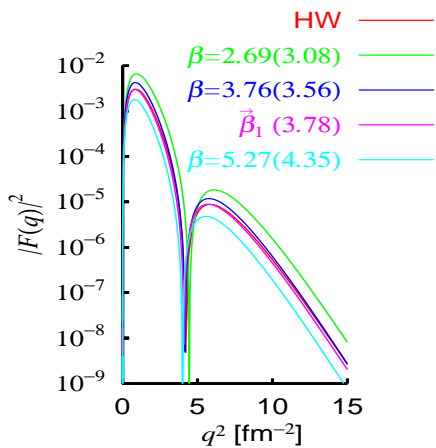}}
\subfigure[]{
\includegraphics[width=6cm,height=6cm]{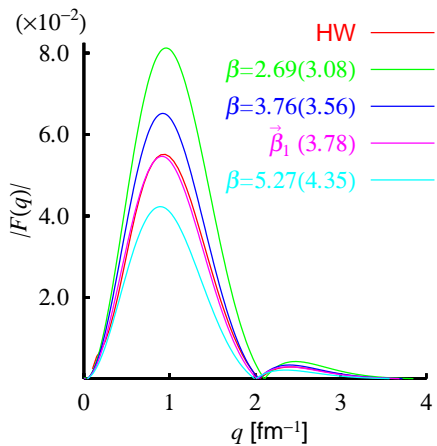}}
\end{center}
\caption{
(Color online) (a): The inelastic form factors to the Hoyle state are plotted as a function of $q^2$. The wave function $\Psi_\perp(\vc{\beta})$ is adopted as the Hoyle state, the size of which is artificially changed by varying the values of the parameter $\vc{\beta}$. $\beta$ is defined as $\beta \equiv \beta_x=\beta_y=\beta_z$ and $\vc{\beta}_1$ denotes $(\beta_x=\beta_y,\beta_z)=(5.27\ {\rm fm},1.37\ {\rm fm}$)~\cite{beta}. The result using the wave functions of ground and Hoyle states which are obtained by solving the Hill-Wheeler equation is denoted by HW. (b): The form factors shown in (a) are replotted as a function of $q$ in linear scale for ordinate. The r.m.s. radii corresponding to $\Psi_{\perp}(\beta)$ are shown in parentheses. Units of all numbers are in fm.}\label{fig:2}
\end{figure}

\begin{figure}[htbp]
\begin{center}
\includegraphics[scale=0.65]{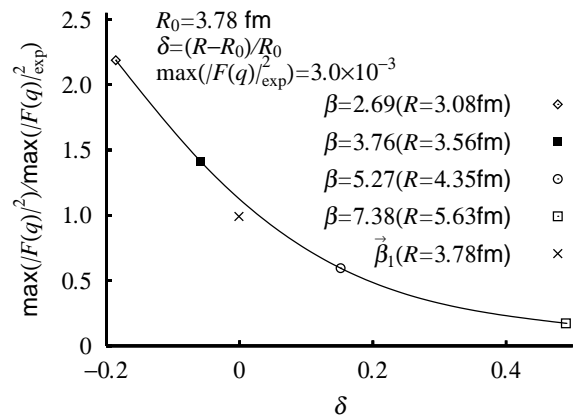}
\end{center}
\caption{
The ratio of the value of maximum height, theory versus experiment, for the inelastic form factor, i.e. max$|F(q)|^2$$/$max$|F(q)|_{\rm exp}^2$, is plotted as a function of $\delta$, which is defined as $\delta=(R-R_0)/R_0$. $R$ and $R_0$ are the r.m.s. radii corresponding to $\Psi_\perp(\beta=\beta_x=\beta_y=\beta_z)$ and $\Psi_\perp(\vc{\beta}_1)$, respectively. Here $R_0=3.78$ fm. Unit of $\beta$ is in fm.}\label{fig:3}
\end{figure}
Here the parameter $\vc{\beta}$ corresponds to the spatial extention of the alpha condensate. In FIG.~\ref{fig:2}, the form factors from the ground state to the Hoyle state calculated at several $\vc{\beta}=(\beta_x,\beta_y,\beta_z)$ values are shown. Short-hand notations $\beta=\beta_x=\beta_y=\beta_z$ and $\vc{\beta}_1 := (\beta_x=\beta_y,\beta_z)=(5.27\ {\rm fm},1.37\ {\rm fm})$~\cite{beta} are used here and in the following. The corresponding r.m.s. radii are shown in parentheses. We see that the overall structure of the form factor is not very much affected by artificially changing the radius of the Hoyle state, i.e. for instance the position of the minimum is only changed in very slight proportions. However, we can see that the amplitude decreases as the r.m.s radius of the $0_2^+$ state increases. In order to see how sensitively the height of the first maximum depends on the r.m.s. radius of the $0_2^+$ state, we plot on FIG.~\ref{fig:3} the variation of this height with respect to the size of the Hoyle state and we see that this height changes strongly when the r.m.s radius of the $0_2^+$ state is changed. It is therefore allowed to say that the measurement of the inelastic form factor of $\alpha$-particle condensate states allows via our model wave function to deduce the radius of such a state. We should note that the wave function of the $0_2^+$ state $\Psi_{\lambda=2}^{J=0}$ can be described rather well by $\Psi_\perp(\vc{\beta})$ as far as reasonable $\vc{\beta}$ values are adopted. The squared overlap between $\Psi_{\lambda=2}^{J=0}$ and $\Psi_\perp(\vc{\beta})$ amounts to $99.2$\% at $\vc{\beta}=\vc{\beta}_1$. Due to this almost complete equivalence between both wave functions, $\Psi_{\lambda=2}^{J=0}$ and $\Psi_\perp(\vc{\beta}_1)$, we understand that the corresponding inelastic form factors obtained by using both wave functions, i.e. denoted by HW and $\vc{\beta}_1$ in FIG.~\ref{fig:2}, almost completely agree with one another. As for the other choices of $\vc{\beta}$, $\Psi_\perp(\vc{\beta})$ also has a reasonably large amount of squared overlap with $\Psi_{\lambda=2}^{J=0}$, i.e. $64.4$\%, $90.1$\%, and $81.8$\% at $\beta=2.69$ fm, 3.76 fm, and 5.27 fm, respectively. It should be emphasized that the fact that the wave function $\Psi_\perp(\vc{\beta})$ which is parametrized by $\beta$ is more or less a good approximation of the $0_2^+$ state guarantees the validity of the above discussion of size dependence.

\begin{figure}[htbp]
\begin{center}
\includegraphics[width=8cm,height=7cm]{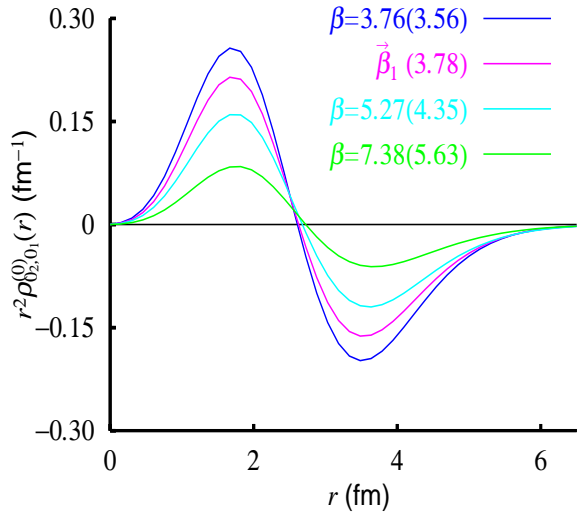}
\end{center}
\caption{
(Color online) Transition densities defined in Eq.~(\ref{eq:2}) multiplied by $r^2$, i.e. $r^2\rho^{(0)}_{0_2 0_1}(r)$, corresponding to wave functions $\Psi_{\perp}(\vc{\beta})$ with different $\vc{\beta}(=\beta_x=\beta_y=\beta_z)$ values. $\vc{\beta}_1$ is given by $(\beta_x=\beta_y,\beta_z)=(5.27\ {\rm fm},1.37\ {\rm fm}$)~\cite{beta}. The r.m.s. radii corresponding to $\Psi_{\perp}(\beta)$ are shown in parentheses. Units of all numbers are in fm.}\label{fig:3a}
\end{figure}

We can analyze the reason for these features of the form factor in the following simple way. In FIG.~\ref{fig:3a} we show the transition density $r^2\rho^{(0)}_{0_2 0_1}(r)$ for different values of $\beta$. Due to the orthogonality between $\Psi_{\lambda=1}^{J=0}$ and ${\widehat P}_\perp {\widehat \Phi}^{N, J=0}_{3\alpha}(\vc{\beta})$ one has the relation $\int_0^\infty r^2\rho^{(0)}_{0_2 0_1}(r) dr =0$. We note that the position of the node of $r^2\rho^{(0)}_{0_2 0_1}(r)$ at $r\approx 2.5$ fm and the point where this transition density drops approximately to zero, i.e. at $r\approx 6.0$ fm, do not depend on the various values of $\beta$. Also the feature of an approximate odd function around the nodal point holds for all $\beta$-values. It clearly can be concluded that one can approximately write 
\begin{equation}
\rho^{(0)}_{0_2 0_1}(r)\approx f(\beta){\tilde \rho^{(0)}}_{0_2 0_1}(r),
\end{equation}
where $f(\beta)\geq 0$ is independent of $r$ and a decreasing function of $\beta$, whereas ${\tilde \rho^{(0)}}_{0_2 0_1}(r)$ is independent of $\beta$. This means that the form factor $F(q)$ just changes amplitude but not shape when the size of the Hoyle state is varied. This analysis is completely consistent with the features seen in FIG.~\ref{fig:2}.

\begin{figure}[htbp]
\begin{center}
\subfigure[]{
\includegraphics[width=6cm,height=6cm]{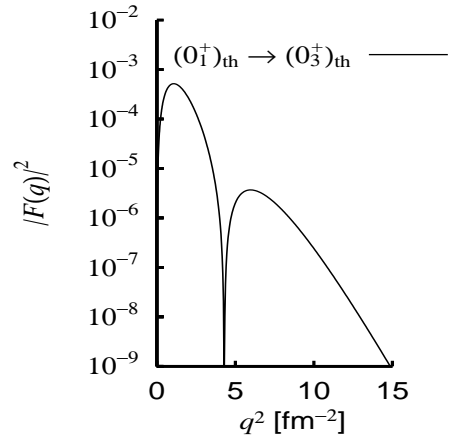}}
\subfigure[]{
\includegraphics[width=6cm,height=6cm]{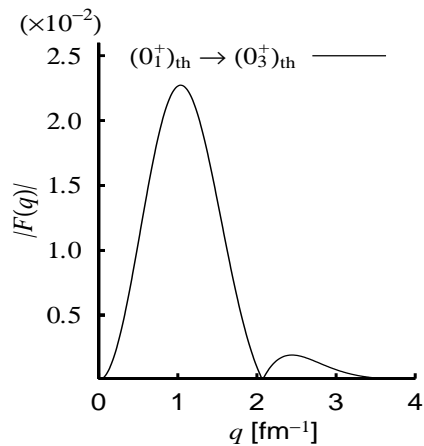}}
\end{center}
\caption{
(a): Inelastic form factor in $^{16}$O to the $4\alpha$ condensate state as obtained from the wave function corresponding to the third eigen energy state in Ref.~\cite{THSR}. (b): The form factor shown in (a) is plotted as a function of $q$ in linear scale for ordinate.}\label{fig:4}
\end{figure}

 On these grounds we also want to make a prediction of the inelastic form factor to the $\alpha$-condensate state in $^{16}$O. In FIG.~\ref{fig:4} we show this form factor calculated with the $\alpha$-particle condensate wave function for $^{16}$O determined previously~\cite{THSR}. This latter state is actually the $3$rd $0^+$ state of our calculation whose energy is at $E_{0_3^+}=14.1$ MeV. We see that the inelastic form factor for $^{16}$O resembles very much in its structure the one of $^{12}$C. The positions of minimum and maximum are almost unchanged, while the height of the first maximum is relatively suppressed compared to the case of $^{12}$C. A candidate of the $4\alpha$ condensate state may have been observed at $13.5$ MeV with an alpha decay width of $0.8$ MeV~\cite{wakasa}. This new state is the $5$th $0^+$ state experimentally, corresponding to the $3$rd $0^+$ state of our calculation. An argument that in $^{16}$O the $\alpha$-condensate state is around $13.5$ MeV could go as follows: It is well known that the second $0^+$ state in $^{16}$O at $6.06$ MeV has a structure of an $\alpha$-particle orbiting in an $S$-wave around an $^{12}$C-core~\cite{carbon,horiuchi_ikeda,tan}. Exciting this $^{12}$C-core to the Hoyle state we find the excitation energy $7.65\ {\rm MeV} + 6.06\ {\rm MeV} =13.71$ MeV. Of course, this close agreement may be a pure coincidence and more experimental evidences are needed.

 The experimental determination of the corresponding form factor would be highly welcome and an eventual agreement with our calculated result, we think, a clear indication of the dilute $\alpha$-particle structure of the corresponding $0^+$ state in $^{16}$O. 

In conclusion, we showed that, without adjustable parameters, our proposed condensate wave function for alpha particles~\cite{THSR} nicely reproduces the experimental inelastic form factor $0_1^+ \rightarrow 0_2^+$ in $^{12}$C. Together with its high sensitivity on the magnitude of the Hoyle state with respect to its size and the reproduction of other experimental data~\cite{THSR,ajz}, we believe that the almost ideal Bose condensate nature of the Hoyle state is now firmly established. We also made predictions for the inelastic form factor to the $2_2^+$ state in $^{12}$C which we interpreted in \cite{funaki2} as a quadrupole particle-hole excitation of the Hoyle state. A prediction of the form factor $0_1^+ \rightarrow \alpha$ condensate state in $^{16}$O is also presented and it is argued that an experimental confirmation of this form factor undoubtedly would reveal the condensate character of the corresponding state.

\section*{Acknowledgements}
This work was partially performed in the Research Project for Study of Unstable Nuclei from Nuclear Cluster Aspects sponsored by Institute of Physical and Chemical Research (RIKEN), and is supported by the Ministry of Education, Culture, Sports, Science and Technology (MEXT), Grant-in-Aid for JSPS Fellows, (No.~03J05511) and by the Grant-in-Aid for the 21st Century COE ``Center for Diversity and Universality in Physics'' from the MEXT of Japan. This work was done as a part of the Japan-France Research Cooperative Program under CNRS/JSPS bilateral agreement.


\begin{thebibliography}{99}
\bibitem{beta}
The $\vc{\beta}_1$ value corresponds to a position of the minimum in FIG.~2 of \cite{funaki}. This may indicate a strongly oblate minimum in contradiction with the spherical $\beta$-values employed here to vary the size of the Hoyle state. It should be recognized, however, in looking at FIG.~2 of \cite{funaki}, that the minimum at $\vc{\beta}=(\beta_x=\beta_y,\beta_z)=(5.7\ {\rm fm},1.3\ {\rm fm})$ is extremely shallow and connected by an almost flat valley to the spherical point. The squared overlap of both wave functions at the minimum and spherical points on the valley is more than $90$\%. This is also underlined by the fact that the $\vc{\beta}_1$ position (cross) in FIG.~\ref{fig:3} only is very little off the continuous line.
\end{thebibliography}
\end{document}